# Flux dynamics in $NdO_{1-x}F_x FeAs$ bulk sample

D. Di Gioacchino · A. Marcelli · S. Zhang · M. Fratini · N. Poccia · A. Ricci · A. Bianconi

**Abstract** We present data of multi harmonic magneto-dynamic experiments. In particular, we performed ac magnetic susceptibility experiments on layered pnictide-oxide quaternary compound NdOFeAs doped with fluorine. The experiments allow measure the critical temperature and probe the flux dynamic behavior using the third harmonic component of the ac susceptibility of a $NdF_{0.16}FeAsO_{0.84}$ bulk sample as a function of temperature and frequency of the applied ac magnetic fields. Measured signals are connected with the non-linear superconducting flux dynamic behavior and are characterized by a 'flux critical states' sustaining a superconducting critical current. In this framework the irreversibility line that describes the stable superconducting state has been extracted from the onset of the third harmonic signal vs. frequency. Finally we present also the analysis of the flux dynamic dimensionality in the investigated sample.

D. Di Gioacchino
I.N.F.N. National Istitute of Nuclear Physics,
National Laboratory of Frascati, Via E. Fermi 40, 00044 Frascati,
Italy
e-mail: daniele.digioacchino@lnf.infn.it

A. Marcelli
I.N.F.N. National Istitute of Nuclear Physics,
National Laboratory of Frascati, Via E. Fermi 40, 00044 Frascati,
Italy

Shuo ZHANG
National Synchrotron Radiation Laboratory
University of Science and Technology of China
230026 Hefei, China
and
Beijing Synchrotron Radiation Facility
Institute of High Energy Physics Chinese Academy of Sciences
19 Yu Quan Lu, 100049 Beijing, China

M. Fratini
University of Rome "La Sapienza", Department of Phisics,
P.le Aldo Moro, 2 - 00185 Rome, Italy

N. Poccia
University of Rome "La Sapienza", Department of Phisics,
P.le Aldo Moro, 2 - 00185 Rome, Italy

A. Ricci
University of Rome "La Sapienza", Department of Phisics,
P.le Aldo Moro, 2 - 00185 Rome, Italy

A. Bianconi
University of Rome, "La Sapienza", Department of Phisics,
P.le Aldo Moro, 2 - 00185 Rome, Italy

## 1 Introduction

The recent discovery of superconductivity in the $LaO_{1-x}F_xFeAs$ system with a transition temperature of 26 K [1] generated a strong interest in the scientific community see Ref.2 and references therein. Besides the relatively high transition temperature, this system that can not be considered analogous to High Temperature Superconducting (HTcSC) cuprate, displays many interesting and unique properties [2]. In particular replacing La with Nd a significant increase of the critical temperature (~ 52 K) has been observed [3]. In addition a structural phase transition centered at 137 K was observed in the NdOFeAs [4,5,6], very similar to the phase transition detected at 165 K in the LaOFeAs [7]. Studies of the flux dynamic are important to characterize and analyze the superconductivity mechanism in cuprates and also in these new materials [8]. Indeed, the flux dynamic response is determined by extrinsic properties e.g., by the pinning effects due to the quenched disorder present inside the material. Moreover, the sample behavior is dominated by the flux dimensionality response with respect to temperature, applied magnetic field and electrical current induced in the sample [9].

The a.c. multi-harmonic susceptibility investigation, in particular the third harmonic ($\chi_3 = \chi_3' + i\chi_3''$), is an effective tool for flux dynamic studies [10]. Actually, the high harmonic signals identify a non-linear response in the flux dynamics, i.e., the flux redistribution in a superconducting state during the ac cycles is dominated by an irreversible state. The boundary between non-linear superconducting processes and linear normal losses processes is named the Irreversible line (IL). In this contribution we present an IL study of the $NdF_{0.16}FeAsO_{0.84}$ bulk sample at low magnetic amplitude vs. frequency. Data are deduced from the onset of the $\chi_3$ measurements in the frequency range 107-1070 Hz and in the temperature range 4-50 K. Moreover, a polar plot $\chi_3'$ versus $\chi_3''$ at 1070 Hz describing the irreversible regime behavior of the $NdF_{0.16}FeAsO_{0.84}$ is compared with the clear and well known 2D (bi-dimensional) behavior of the Bi2223.

## 2 Experimental

The superconducting $NdF_xFeAsO_{1-x}$ bulk sample has been prepared by high pressure synthesis from Nd, As, Fe, $Fe_2O_3$, $FeF_3$ powders (the purity of all starting chemicals was better than 99.99%). The components were mixed together according to the nominal stoichiometric ratio of $Nd[O_{0.84}F_{0.16}]FeAs$, then grounded thoroughly and pressed into small pellets. The latter were sealed in boron nitride crucibles and sintered in a high-pressure synthesis apparatus under a pressure of 6 GPa and at the temperature of 1300 ºC for two hours [11]. The sample dimensions are 1.5 x 1.5 x 1.0 mm. The ac high harmonics susceptibility response of the sample has been measured in a susceptometer based on pick-up double coils surrounded by the driven coil (fig.1) [12]. The sample is typically set on a sapphire holder, which fits in the pick-up coil bridge. Two coils connected in series and rolled up in opposition realize the bridge and while one is strongly coupled (pick-up coil), the second one is weakly (balance coil) coupled to the sample. The sample is cooled in the 'Zero magnetic Field Cooling (ZFC) set-up' in a thermally controlled He gas flow cryostat. The ac driving magnetic field frequency range chosen is 107<f<1070 Hz with amplitude of 1.3 Gauss. A multi harmonic lock-in amplifier measures the induced signal. The sample was oriented parallel to the ac magnetic field, and during the acquisition the temperature was measured by a CGR thermometer. The susceptibility data are the result of the subtractions between the signals collected with and without the sample for each frequency and temperature. The $\chi_3$ onset has been determined by the temperature at which the third harmonic amplitude $\chi_3$ starts to deviate from zero.

## 3 Analysis and discussion

In figure 2, we show the first harmonic real part $(\chi'_1)$ exhibiting a clear superconducting diamagnetic behavior with $T_c$=46.5 K, while the Imaginary part ($\chi''_1$) returns the area of the magnetic cycle determined by the sample under the excitation of the ac magnetic field. In this plot both hysteretic and normal losses behaviors are observed. Hysteretic losses describe the interaction between magnetic fluxes and pinning that sustain the superconductivity and in particular the critical current. Normal losses, such as single electron and flux flow motions are the processes that limit the superconductivity in the sample. In figures 3 and 4 we show both the Real and Imaginary parts of the third harmonic ($\chi'_3$, $\chi''_3$) vs. temperature and frequency. These behaviors are connected only with non-linear flux dynamic and hysteretic features i.e., the high harmonic ac susceptibility probes only the flux-pinning interaction. Both figures 3 and 4 showed oscillating signals around the zero value. Third harmonic amplitudes are connected with the knee curvature of the non-linear I-V superconducting characteristic around the critical current value [13]. In figure 3, $\chi'_3$ frequency dependence in the range 40 - 45 K shows a clear change of the I-V knee curvature at different frequencies for the same temperature, e.g. increasing the frequency we observed an increase of the critical current induced in the sample. Moreover, $\chi_3$ oscillating behaviors show that the susceptibility harmonic components are polynomial functions of the ratio between magnetic applied field and induced magnetic field connected with the critical current induced in the sample [14].

Looking at both figures 3 and 4 we have also to underline that $T_{on}(\chi_3)$ shows a very weak frequency dependence. In a typical ZFC procedure, after cooling, a magnetic field is applied to the system which switches in a flux-pinning critical state configuration and subsequently decays via a creep process in a final stable flux-pinning 'Glass state' pattern [5]. In our case, at low magnetic field, in the $NdF_{0.16}FeAsO_{0.84}$ sample the 'critical state' decayed in a 'Glass state' with a faster characteristic creep time rate respect to the experimental time window range (9.3-0.93 msec) used, corresponding to the applied field frequency (107-1070 Hz) of the measurements.

Finally, in the Cole-Cole plot ($\chi'_3$ vs $\chi''_3$) of figure 5 we compare at 1070 Hz and 1.3 G both $NdF_{0.16}FeAsO_{0.84}$ and BiSCCO 2223 samples. From comparison we may clearly recognize that the new $NdF_{0.16}FeAsO_{0.84}$ superconductors exhibits a behavior similar to that of BiSCCO 2223, a well know 2D superconductor in a flux dynamic regime [15]. This result addresses a possible 2D flux dynamic behavior also for this system.

**Acknowledgements:** We gratefully acknowledge also the support of the Italian Ministry Foreign Affairs in the framework of the 12th Executive Programmer of scientific and Technological Cooperation between the Italian Republic and the People's Republic of China.

**Figure captions**

Fig.1 - The a.c. multi-harmonic susceptometer installed in the He gas temperature variable cryostat (4-300K) of the Laboratori Nazionali di Frascati of the INFN

Fig.2 – The Real and Imaginary parts of first harmonic susceptibility ($\chi'_1$, $\chi''_1$) vs. temperature at f=107 Hz and at |Hac|=1.3 Gauss

Fig.3 – The Real part of the third harmonic susceptibility ($\chi'_3$) vs. temperature in the range 107-1070 Hz and at $|H_{ac}|$=1.3 Gauss

Fig.4 – The Imaginary part of the third harmonic susceptibility ($\chi''_3$) vs. temperature in the range 107-1070 Hz and at $|H_{ac}|$=1.3 Gauss

Fig.5 – The Cole-Cole plot ($\chi'_3$ vs. $\chi''_3$) of $NdF_{0.16}FeAsO_{0.84}$ (red) and of BiSCCO 2223 (blue) at 1070 Hz and at $|H_{ac}|$=1.3 Gauss

FIGURE 1

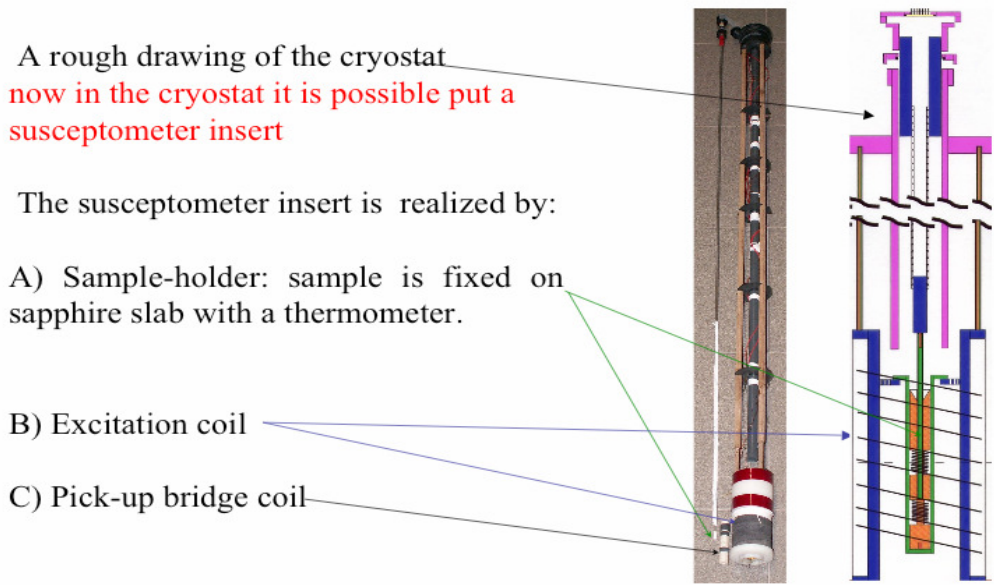

The susceptibility set-up

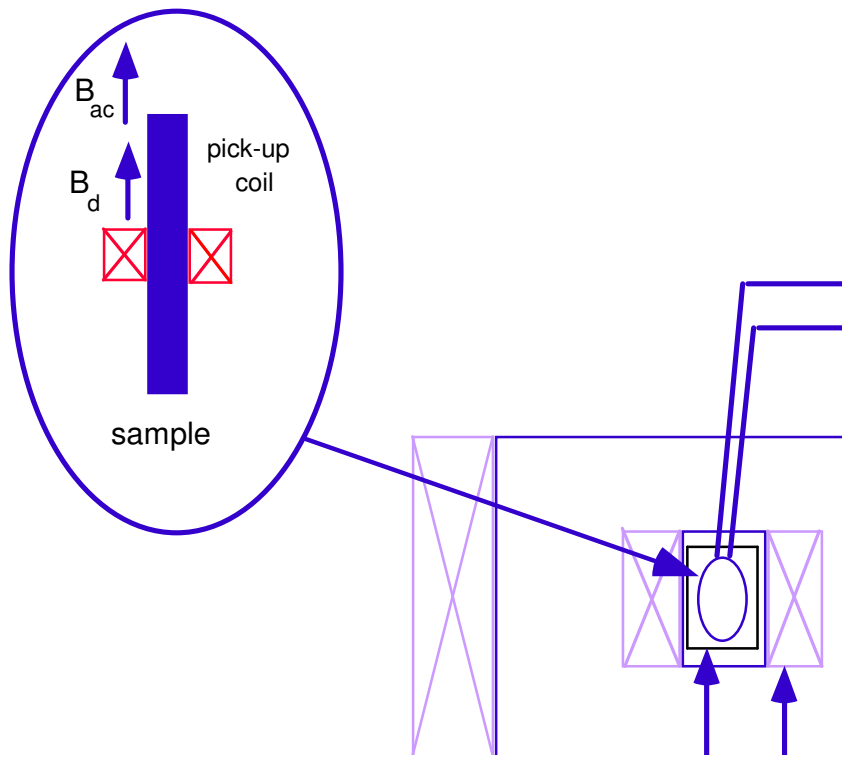

FIGURE 2

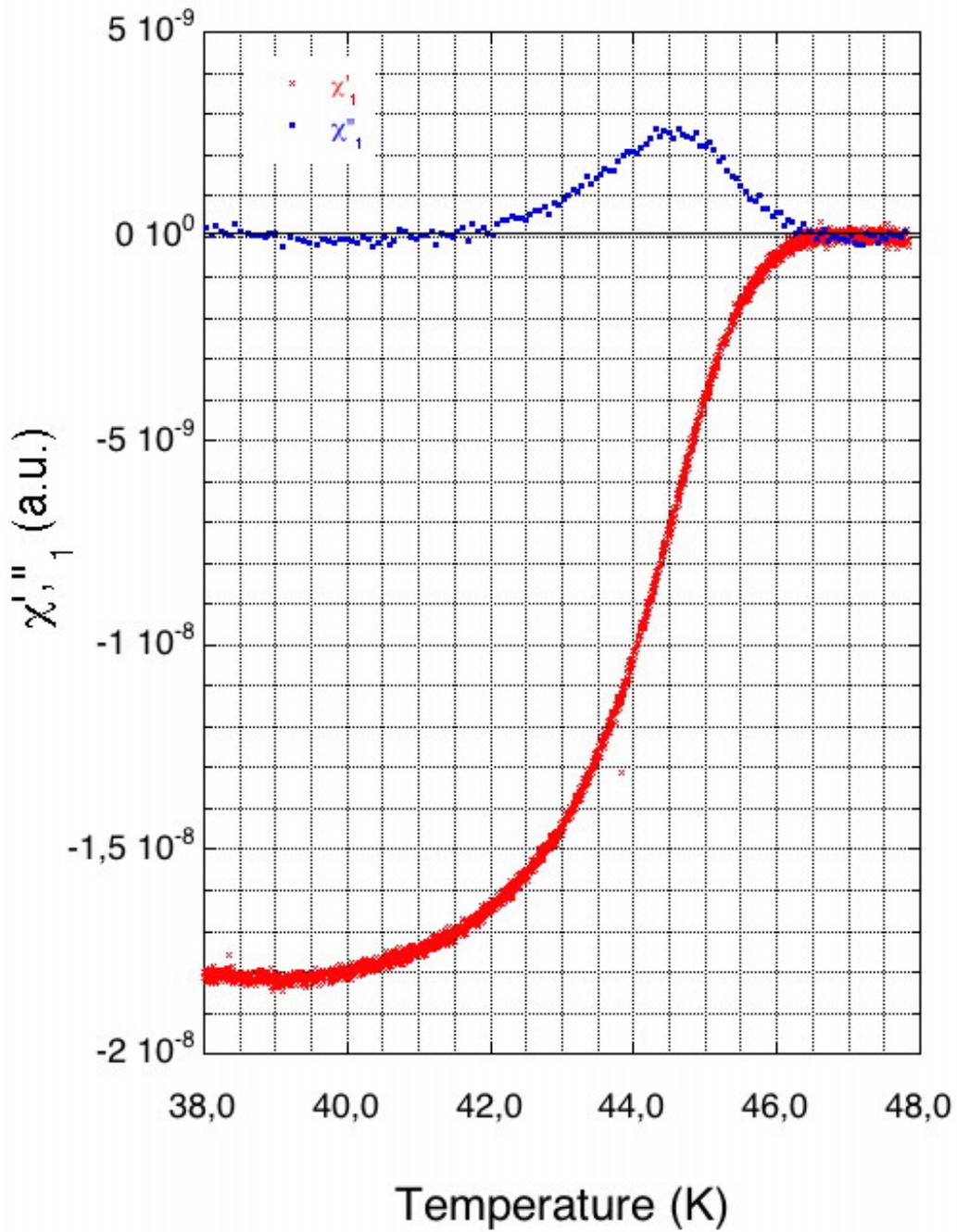

FIGURE 3

Figure 3: Plot of $X'_3$ (a.u.) versus Temperature (K) at frequencies 107Hz, 207Hz, 307Hz, 507Hz, 707Hz, and 1070Hz, with $|H_{ac}|=1.3$ Gauss.

FIGURE 4

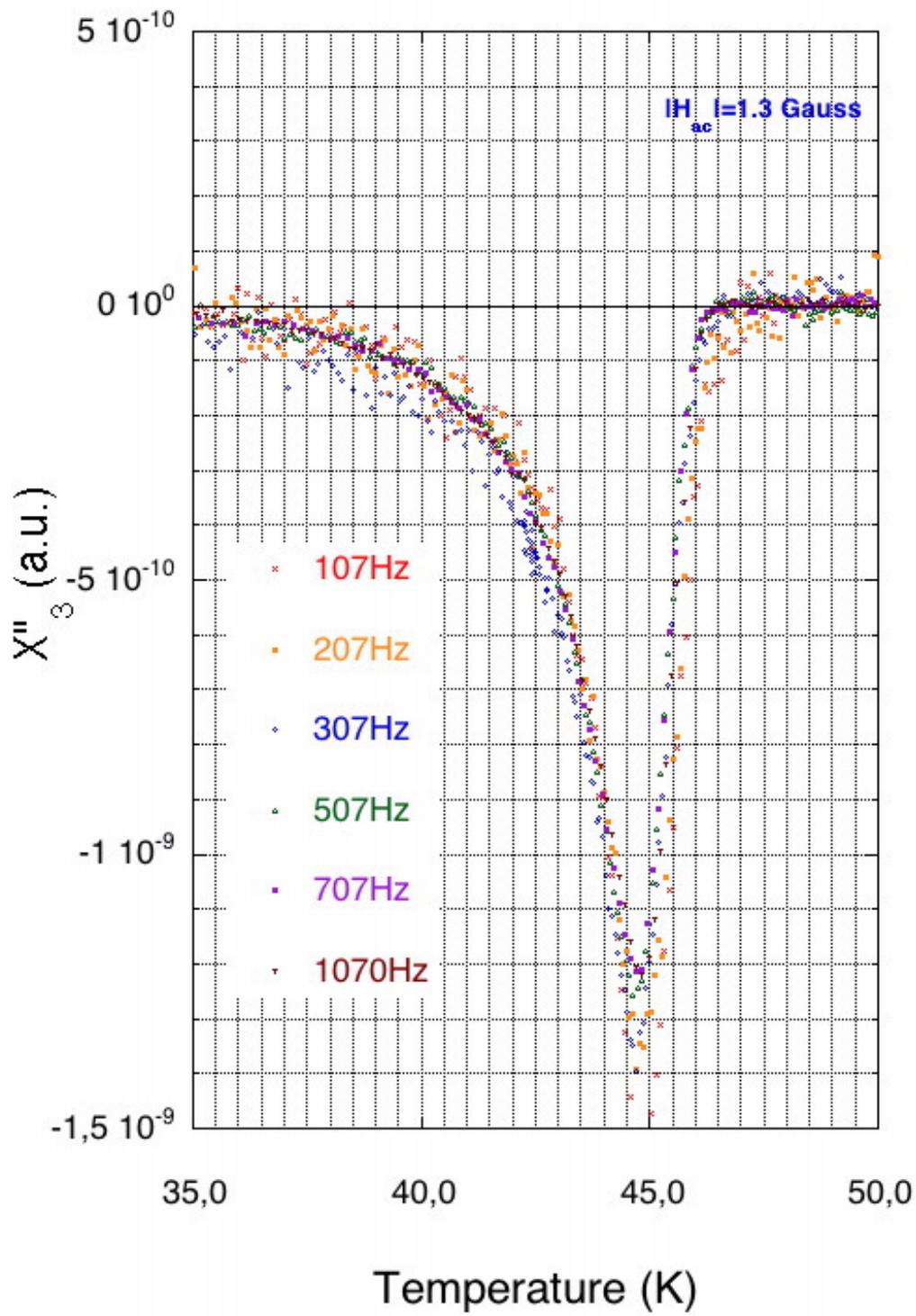

FIGURE 5

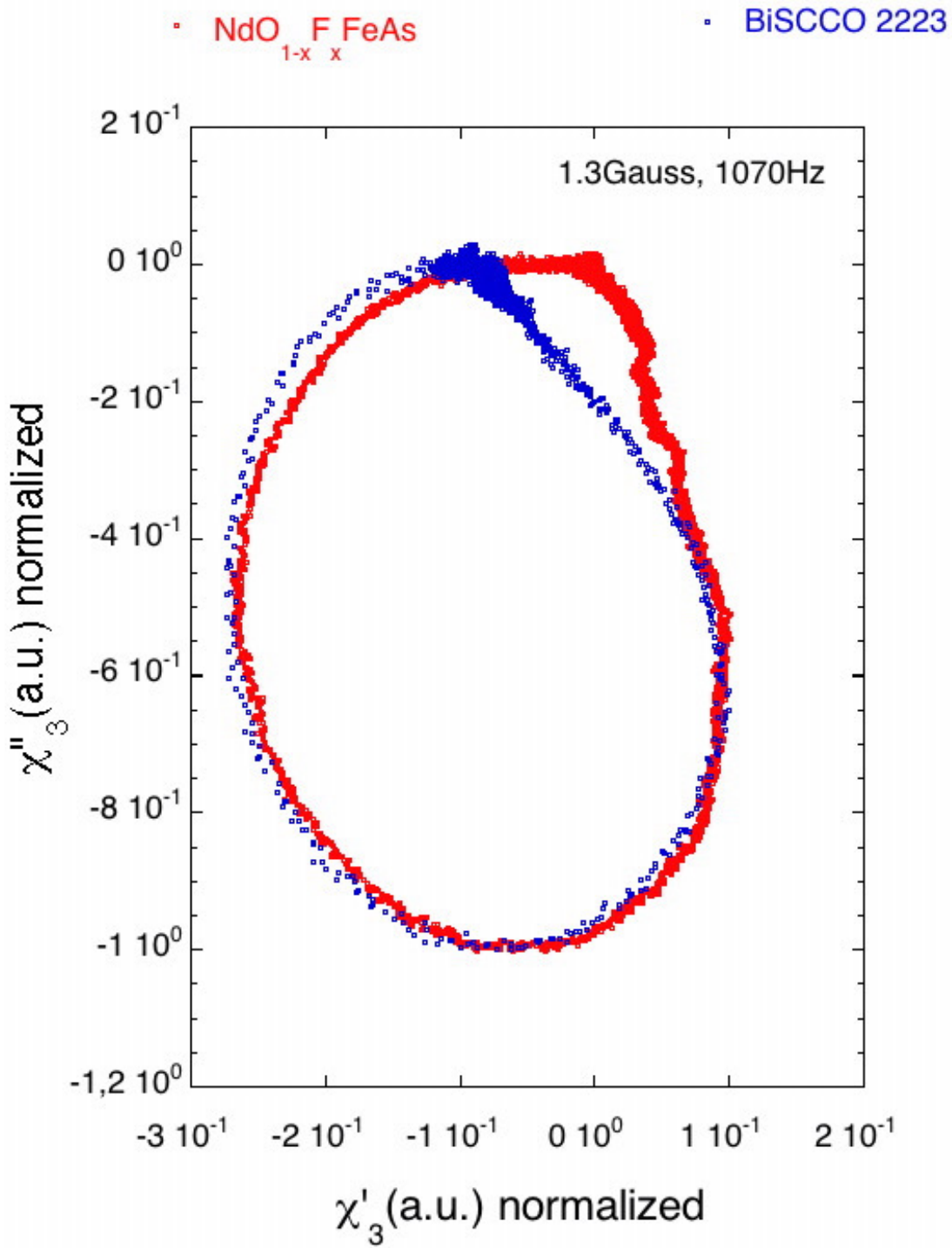